\documentclass[graybox]{svmult}

\usepackage{mathptmx}       
\usepackage{helvet}         
\usepackage{courier}        
\usepackage{type1cm}        
\usepackage{makeidx}         
\usepackage{graphicx}        
\usepackage{multicol}        
\usepackage[bottom]{footmisc}

\makeindex             

\begin{document}

\title*{Rings and bars: unmasking secular evolution of galaxies}
\author{Johan H. Knapen}
\institute{Johan H. Knapen \at Instituto de Astrof\'\i sica de Canarias, E-38200 La Laguna, Tenerife, Spain and 
Departamento de Astrof\'\i sica, Universidad de La Laguna, E-38205 La Laguna, Tenerife, Spain, \email {jhk@iac.es}}
%
%
\maketitle

\abstract{Secular evolution gradually shapes galaxies by internal processes, in contrast to early cosmological evolution which is more rapid. An important driver of secular evolution is the flow of gas from the disk into the central regions, often under the influence of a bar. In this paper, we review several new observational results on bars and nuclear rings in galaxies. They show that these components are intimately linked to each other, and to the properties of their host galaxy. We briefly discuss how upcoming observations, e.g., imaging from the {\it Spitzer} Survey of Stellar Structure in Galaxies (S$^4$G), will lead to significant further advances in this area of research.}

\section{Introduction}
\label{intro}

The general topic of galaxy evolution is enjoying widespread attention in the literature. The aim is to answer, through observations and numerical modelling, how galaxies have evolved from the earliest stages of the Universe, or from a subsequent epoch of formation, to the shape in which we observe them in the local Universe. The two main strands in this wide topic are cosmological evolution of galaxies, which deals with their formation and early evolution, and secular evolution, with which we mean the internal evolution of galaxies, under the influence of the dynamical actions of, e.g., bars or spiral arms. Secular evolution, as comprehensively reviewed by Kormendy \& Kennicutt (2004),  is a relatively slow process compared to the more rapid evolution undergone by galaxies in the early Universe, the latter often due to mergers and galaxy-galaxy interactions.

Because the early stages of galaxy formation and evolution are hard or sometimes impossible to observe due to the combined effects of distance, redshift, and dust extinction, the detailed study of nearby galaxies is one of the very few ways to confirm the detailed predictions of models of large-scale galaxy formation and evolution. This kind of study can be nicknamed `galactic palaeontology', because in local galaxies we study the `fossil record' of billions of years of galaxy evolution---both cosmological and secular. To read and interpret this fossil record, a combination of many different observational and interpretational techniques must be used, from observations of individual stars in our own Milky Way and the most nearby galaxies to photometric and kinematic observations across many different wavelengths in external galaxies, all combined with a wide range of interpretational, analytical and numerical tools.

One of the major drivers of the internal evolution is the flow of gaseous material, from the disk to the central regions of the galaxy. To move inwards, rotating gas must lose angular momentum, and it can do so by shocking and under the influence of a non-axisymmetric potential. That in turn gets set up by a bar, by interactions or minor mergers, or even by less obvious deviations from axisymmetry like spiral arms, ovals, or lenses (e.g., Schwartz 1984; Shlosman et al. 1989, 1990; Knapen et al. 1995; Kormendy \& Kennicutt 2004; Comer\'on et al. 2010). Bars are very common in galaxies, about two thirds of local galaxies have a bar (de Vaucouleurs et al. 1991; Sellwood \& Wilkinson 1993; Moles et al. 1995; Ho et al. 1997; Mulchaey \& Regan 1997; Hunt \& Malkan 1999; Knapen et al. 2000; Eskridge et al. 2000; Laine et al. 2002; Laurikainen et al. 2004a; Men\'endez-Delmestre et al. 2007; Marinova \& Jogee 2007; Sheth et al. 2008; Laurikainen et al. 2009). As we will see below, in Sect.~\ref{rings}, nuclear rings can occur in unbarred galaxies, and they seem to prove that the non-axisymmetry induced by a spiral or oval may well be enough to induce gas inflow, and thus lead to secular evolution. Although most galaxies are barred, secular evolution is thus not dependent on the presence of a bar.

Galactic interactions are rare, occurring in only about 2\% of local galaxies (up to 4\% if merely bright galaxies are considered; Knapen \& James 2009). Close companions to local galaxies are much more common, and Knapen \& James (2009) found that some 15\% of local galaxies have a companion not more than 3\,mag fainter than itself within a radius of five times the diameter of the galaxy under consideration, and within a range of $\pm200$\,km\,s$^{-1}$ in systemic velocity. The effects on the star formation rate in galaxies of the presence of a close companion, and even of interactions, is, perhaps surprisingly, limited (observations by Bushouse 1987; Smith et al. 2007; Woods \& Geller 2007; Li et al. 2008; Knapen \& James 2009; Jogee et al. 2009; Rogers et al. 2009; Ellison et al. 2010, corroborated by numerical simulations by Mihos \& Hernquist 1996; Kapferer et al. 2005; Di Matteo et al. 2007, 2008; Cox et al. 2008). Statistically, the star formation rate is raised by a factor of just under two by the presence of a close companion, but the H$\alpha$ equivalent width is hardly increased at all (Knapen \& James 2009). This implies that even though galaxies with close companions tend to form stars at a higher rate, they do so over extended periods of time, and not as a burst. Even the majority of the truly interacting galaxies in Knapen \& James's (2009) sample have unremarkable star formation properties. The reason that extreme star-forming galaxies, such as ultra-luminous infrared galaxies (ULIRGs) are found to be almost ubiquitously interacting must surely be a selection effect. In general, interactions do not always cause starbursts, and starbursts do not always occur in interacting galaxies.

In fact, Knapen \& James (2009) used their sample of 327 nearby disk galaxies to explore how one might best define the term `starburst'. They concluded that {\it none} of the definitions that are in common use in the literature can be considered to be objective and generally discriminant. For instance, selecting galaxies on the basis of their high star formation rate yields large star-forming disk galaxies. Selecting those with high equivalent widths (apparently a {\it bona fide} starburst discriminator as this selects galaxies with a much enhanced current star formation rate as compared to the average rate in the past), yields primarily late-type galaxies of very small mass, whose star formation activity is caused by one or a few H{\sc ii} regions (and which will have very low impact on the intergalactic medium through, e.g., stellar winds). And selecting galaxies with the shortest gas depletion timescales does not only select galaxies with very high current star formation rates, but also gas-poor early-type galaxies with a very small star formation rate. The conclusion of Knapen \& James (2009) is that starbursts are very hard to define properly, and the use of the term should be restricted to well-described small numbers of objects. 

This review deals with aspects of secular evolution, and how we can trace its actions back through the detailed study of structural components, particularly bars and rings, in nearby galaxies. The interpret the effects of these agents, the tools we will use here are primarily optical and infrared imaging and two-dimensional kinematic mapping. Other authors have presented reviews on galactic evolution, and in particular the paper by Kormendy \& Kennicutt (2004) presents an authoritative review of the subject of secular evolution. We will supplement that by presenting selected recent results on bars and rings that highlight the intricate and detailed connections that exist between the different structural components of a galaxy and the overall galactic evolution. 

In Section~\ref{bars} of this paper, we will review how the strength of a bar is connected to many of the basic properties of the bar, such as its length, or the shape of its dust lanes. We will also see that bars are indeed connected to spirals, and how bars in S0 galaxies may be different from those in spirals. Section~\ref{rings} describes how the basic physical properties of the host galaxy and, where present, its bar condition the location and morphology of a nuclear ring, thus highlighting the close physical connections between these components. Section~\ref{future} discusses how the {\it Spitzer} Survey of Stellar Structure in Galaxies (S$^4$G; Sheth et al. 2010) will deliver the data which should allow us to make significant further progress in the study of galaxy evolution by means of detailed analyses of the stellar component in a large sample of nearby galaxies. We briefly present our conclusions in Section~\ref{conclusions}.

\section{Bars}
\label{bars}

Bars are common in galaxies: around one third of all disk galaxies have a strong bar, and when including weak bars and ovals, the fraction of barred galaxies rises to well over two-thirds (references in the introductory section). The bar fraction is roughly invariant with morphological type across the spirals (Knapen et al. 2000), but lower in the S0s (Barazza et al. 2008; Laurikainen et al. 2009). There is an interesting debate in the literature over whether the fraction of bars is lower (Abraham et al. 1999; van den Bergh et  al. 2002; Sheth et al. 2008) or not (Elmegreen et al. 2004; Jogee et al. 2004; Zheng et al. 2005) at the highest redshifts where bars can be reliably recognised, around $z\sim1$. Nair \& Abraham (2010) find that the fraction of barred spiral galaxies between redshifts of 0.01 and 0.1 is a strong function of stellar mass and star formation history. They suggest that the discrepancy in the reported bar fraction evolution with redshift may have its origin in observational biases and selection criteria, because the strong bar fraction is sensitive to the mass range that is being probed (see also Cameron et al. 2010, who find that the early- and late-type bar fractions vary with redshift and galaxy mass).

Bars are efficient agents of angular momentum transfer in galaxies, and are expected to evolve as they transfer angular momentum to the disk and the halo, and allow the radial inflow of gas (e.g., Lynden-Bell \& Kalnajs 1972; Sellwood 1981; Shlosman et al. 1989, 1990; Debattista \& Sellwood 1998; Athanassoula 2002, 2003; Martinez-Valpuesta et al. 2006). As a result, bars should slow down and grow over time, and become longer, thinner, and stronger (Athanassoula 2003). As bars are relatively easy to observe, they can be used as effective probes of secular evolution in galaxies.

\subsection{Basic properties of bars}

\begin{figure}
\includegraphics[width=0.5\textwidth]{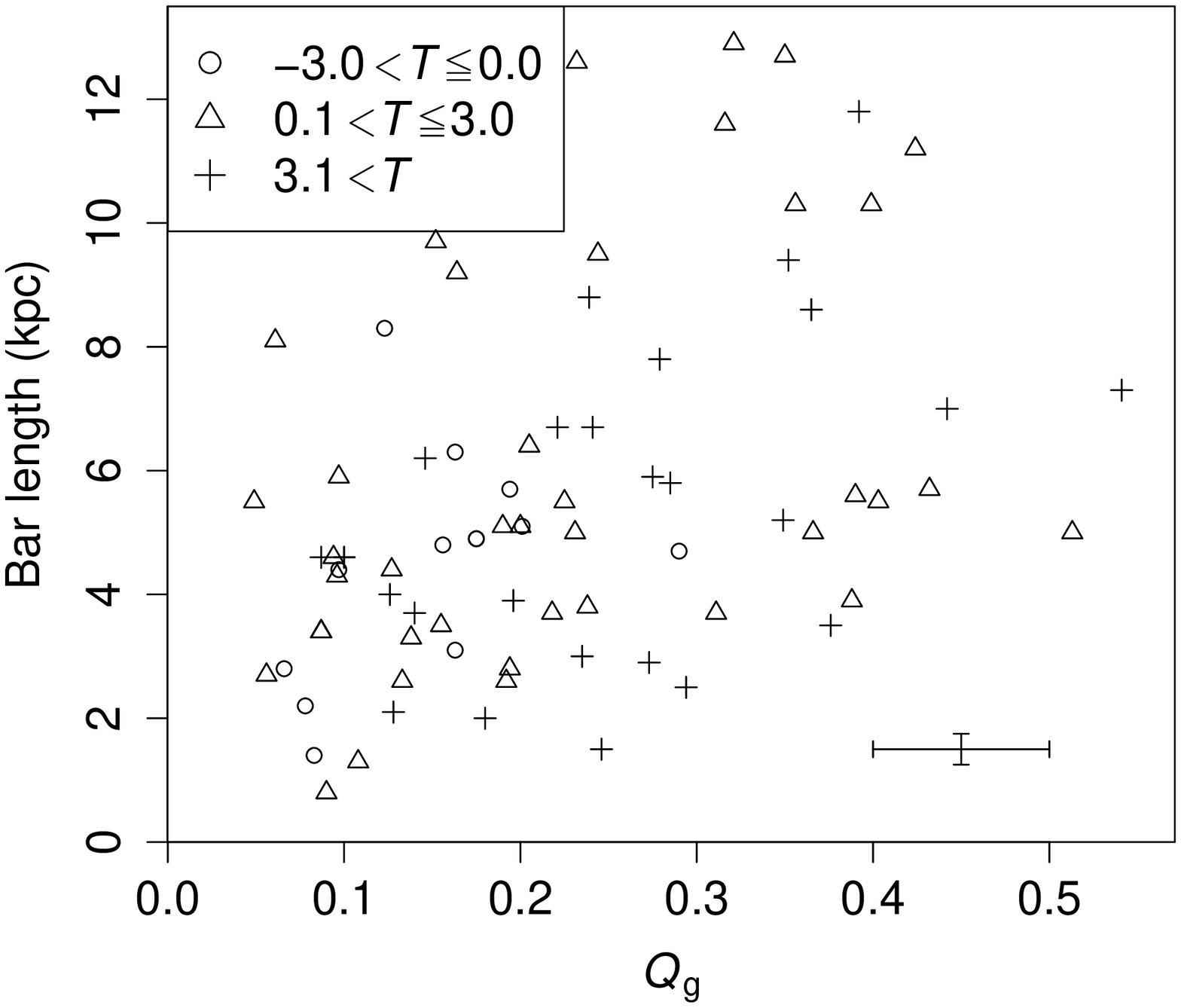}
\includegraphics[width=0.5\textwidth]{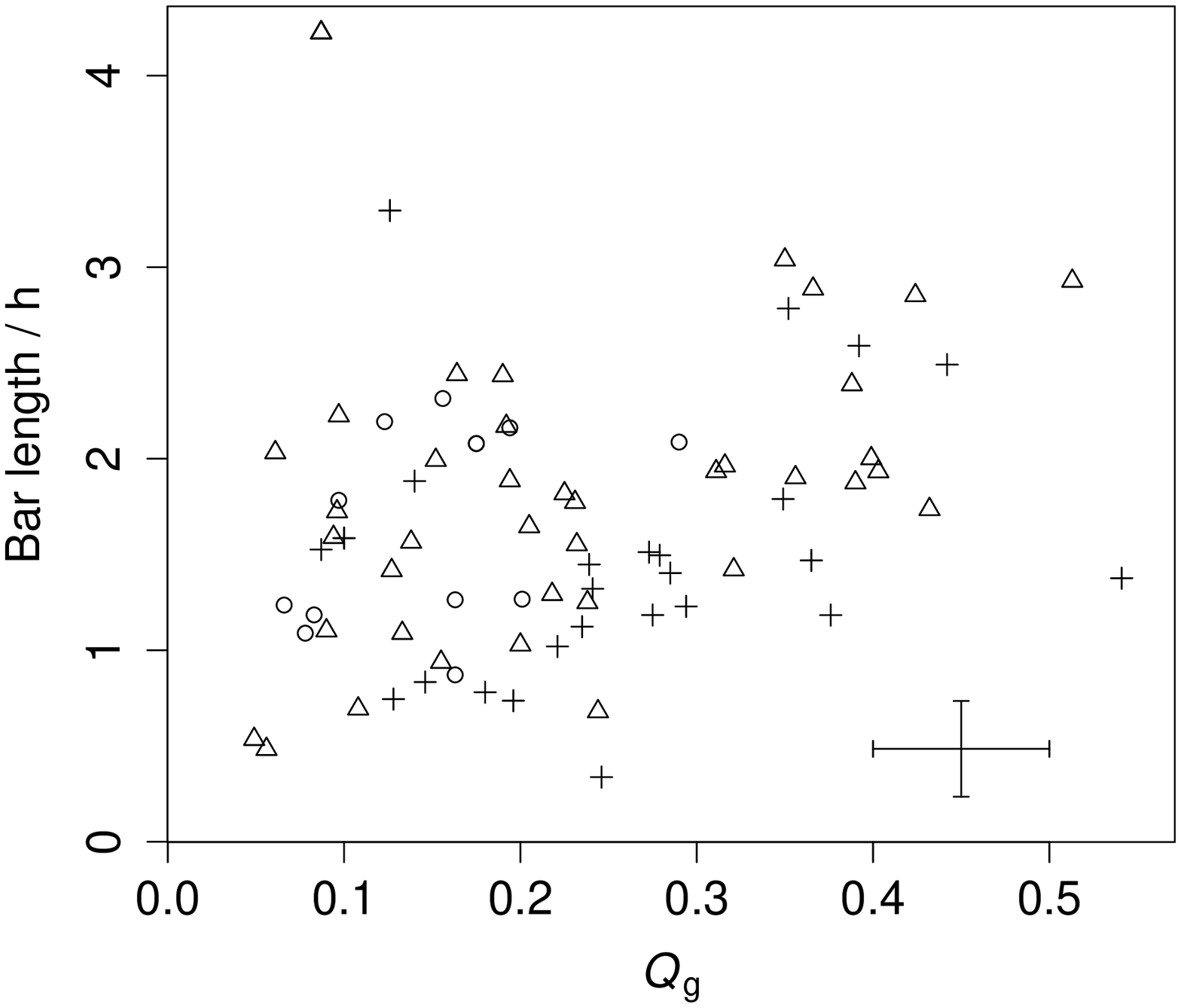}\\
\includegraphics[width=0.5\textwidth]{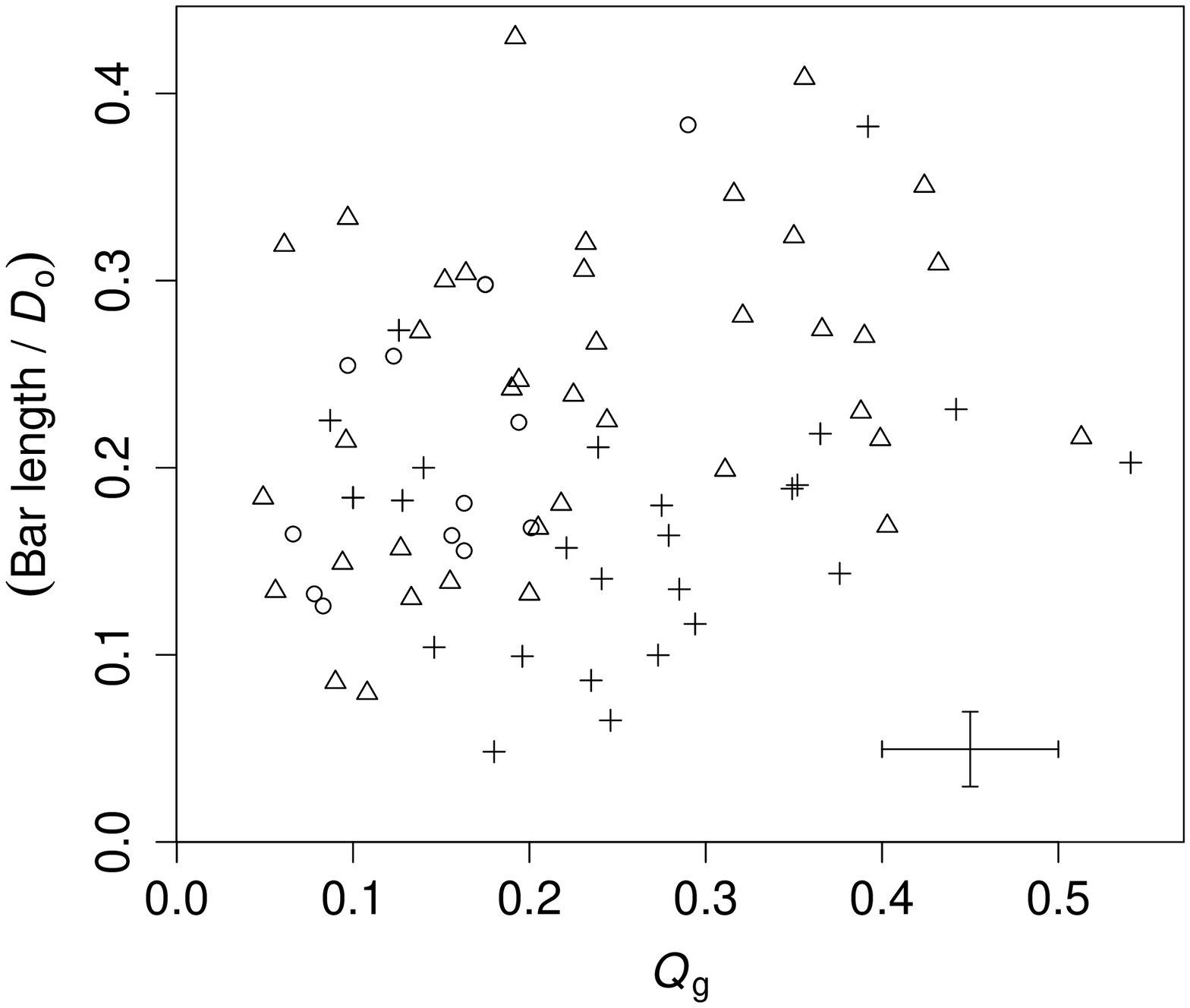}
\includegraphics[width=0.5\textwidth]{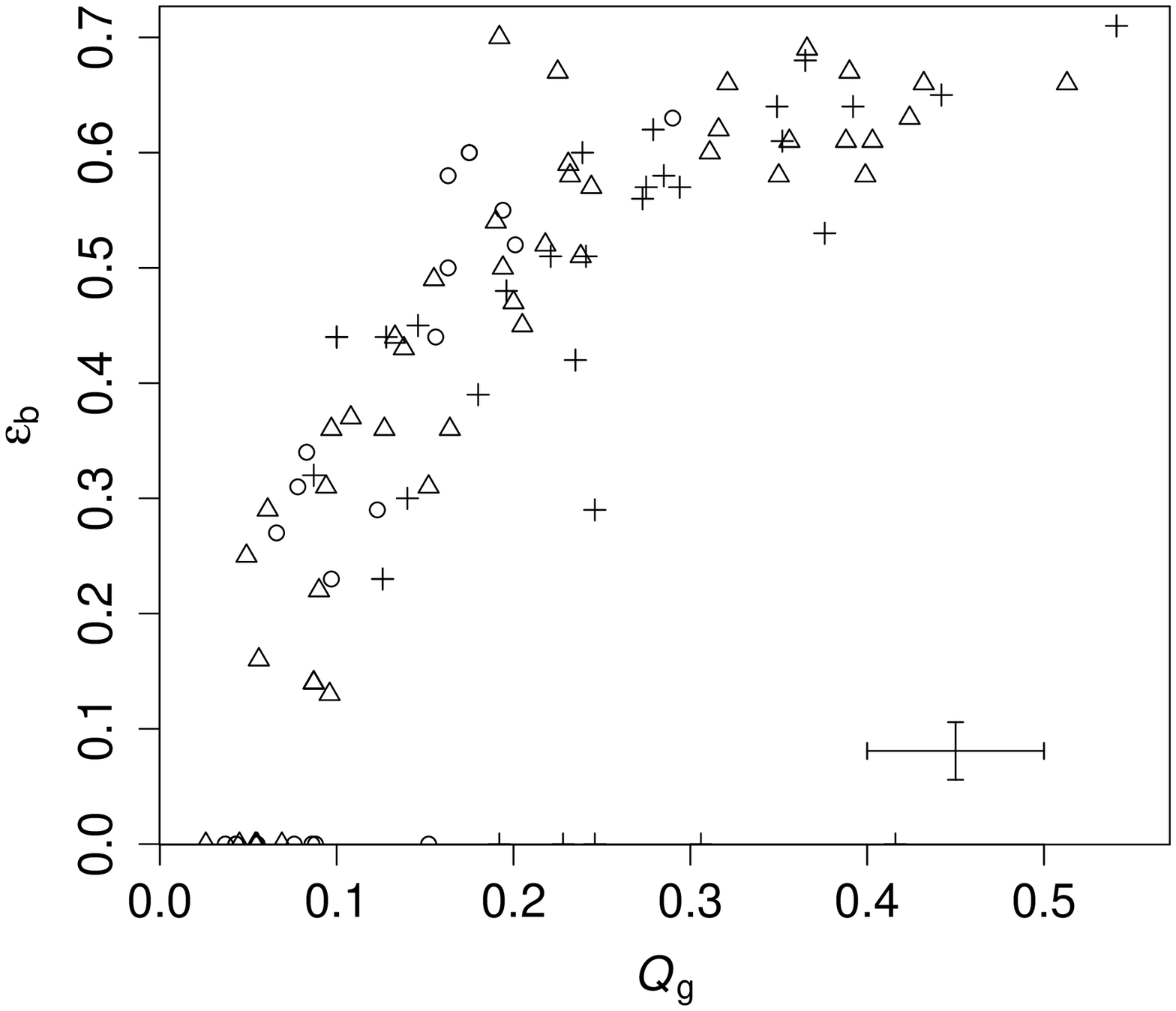}
\caption{Bar length and bar ellipticity (the latter in the lower right panel) as a function of the bar strength parameter $Q_{\rm g}$, for 107 nuclear ring host galaxies. The top left panel shows the absolute bar length, the top right panel the bar length normalised by the scale length of the exponential disk, and the bottom left panel the bar length normalised by the disk size. Typical uncertainties are indicated in each panel, and the symbol styles indicate different morphological types as shown in the top left. Reproduced with permission from Comer\'on et al. (2010).}
\label{AINUR-bars}       
\end{figure}

To study their basic properties and the relations to their host galaxies, bars are parametrized by determining their main parameters, such as length, thickness, and strength. Many different definitions of such parameters can be devised and have been presented in the literature, but modern analyses use techniques such as image decomposition, Fourier analysis, and the determination of bar strength parameters from deep high-quality images. This is done preferably in the infrared where light traces mass better than in the optical, and where the effects of dust extinction are greatly reduced. As an illustration of the kind of relations that are being found, we show in Fig.~\ref{AINUR-bars} the measurements that have been obtained by Comer\'on et al. (2010) as part of their Atlas of Images of Nuclear Rings (AINUR), of which we will later review more results on nuclear rings. This is not an unbiased survey, but its results do illustrate the points we wish to highlight here.

Fig.~\ref{AINUR-bars} shows various measures of the length of the bar, as well as ellipticity, as a function of the bar strength parameter $Q_{\rm g}$, and indicating three ranges of morphological types. $Q_{\rm g}$ is a non-axisymmetric torque parameter, quantifying the impact of non-axisymmetries in a galaxy by measuring the maximum value of the tangential forces normalized by the axisymmetric force field (e.g., Combes \& Sanders 1981; Quillen et al. 1994; Salo et al. 1999; Block et al. 2001; Buta \& Block 2001; Laurikainen \& Salo 2002)\footnote{Note that $Q_{\rm g}$ is measured across a whole galaxy and may include contributions from the spiral arms as well as the bar. In earlier work, what we now refer to as $Q_{\rm g}$ was called $Q_{\rm b}$. In more modern work, we refer to $Q_{\rm b}$ as the non-axisymmetric torque from the bar only, after separating the contributions of bar and spiral arms (e.g., Buta et al. 2006). In early-type galaxies, including S0s, the spiral arm contribution is very small and $Q_{\rm g}$ can be used as $Q_{\rm b}$.}. Comer\'on et al. (2010) determined  $Q_{\rm g}$ values using azimuthal Fourier decomposition and polar integration (following Salo et al. 1999 and Laurikainen \& Salo 2002). 

The data points are from a sample of 107 galaxies hosting nuclear rings (Comer\'on et al. 2010). They confirm the general trend, already noted by Martinet \& Friedli (1997) and Laurikainen et al. (2004b), among others, for bars to become longer as they become stronger, but the scatter is rather large. This conclusion can be drawn from any of the three bar length measures plotted, directly in units of kpc, and normalised by the disk scale length $h$ or the disk size $D_0$. Two regions of the various diagrams in Fig.~\ref{AINUR-bars} are empty, which seems to indicate that there are no very strong short bars, nor any very weak and long ones. In fact, some late-type galaxies do have short bars with large $Q_{\rm g}$, due to the fact that they have very little, or no, bulge to dilute the non-axisymmetric contribution of the bar to the potential (e.g., Laurikainen et al. 2004b). This effect also explains why $Q_{\rm g}$ decreases but the bar length increases towards the earlier Hubble types (Buta et al. 2005; Laurikainen et al. 2007; Fig.~\ref{AINUR-bars}). The bar ellipticity is seen to increase monotonically with $Q_{\rm b}$ (Buta et al. 2004), with some scatter but very few outliers. 

From deep $K_{\rm s}$-band images of 20 galaxies, Elmegreen et al. (2007) confirmed independently that longer bars are stronger (as indicated in their paper by the peak amplitude of the normalised $m=2$ Fourier component). They also found an interesting correlation between the bar length and an increased density in the central parts  of the disks. Elmegreen et al. note that, as dense galaxies evolve faster, these results indicate that bars grow in length and amplitude, with the densest galaxies showing the fastest evolution. Numerical modelling provides theoretical support for correlations such as the one between bar strength and length (see, e.g., the recent simulations by Villa-Vargas et al. 2010, with varying gas fractions and gas spatial resolution).

Other aspects of bars which are directly related to secular evolution include the frequency of ansae-type morphology, significantly enhanced in early-type galaxies (Laurikainen et al. 2007; Mart\'\i nez-Valpuesta et al. 2007); the finding that bars with double-peaked amplitude profiles are stronger than those with single-peaked profiles (Buta et al. 2006), and occur more frequently in early-type galaxies than in spirals (Laurikainen et al. 2007), which might indicate that they are more evolved (Athanassoula \& Misioritis 2002; Athanassoula 2003); and the fact that most S0s have lenses (Laurikainen et al. 2007).

\subsection{Bar dust lanes}

A further hint at the internal physical processes in bars is provided by the study of dust lanes of Comer\'on et al. (2009). Dust lanes have been recognised in relation to shocks in the flow of gas in barred galaxies now almost 50 years ago (Prendergast 1962), and Athanassoula (1992) in fact predicted from numerical modelling that the degree of curvature of the main dust lanes in a bar should decrease as the bar gets stronger: stronger bars have straighter dust lanes. As dust lanes are clearly visible in barred galaxies (e.g., Pease 1917; Sandage 1961), this is in principle an easily observable tracer of the fundamental physics and dynamics of galactic bars, yet with the exception of the preliminary study by Knapen et al. (2002) of only nine galaxies which confirmed the predictions, it had never been tested observationally. Comer\'on et al. (2009) collected images of 55 bars of which the shape of the dust lanes could be measured, and for which the bar strength $Q_{\rm b}$ was available from the literature. 

\begin{figure}
\includegraphics[width=0.5\textwidth]{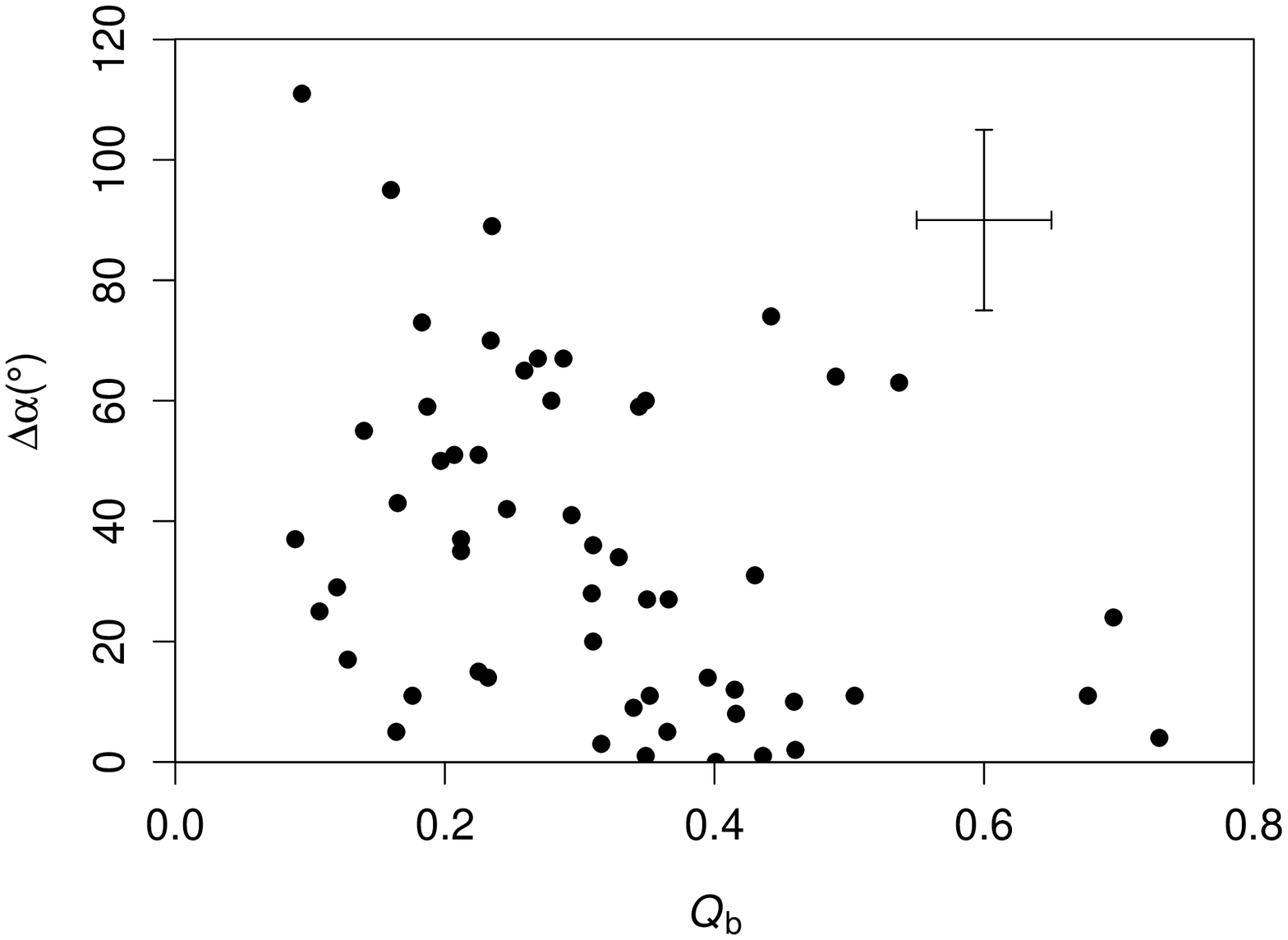}
\includegraphics[width=0.5\textwidth]{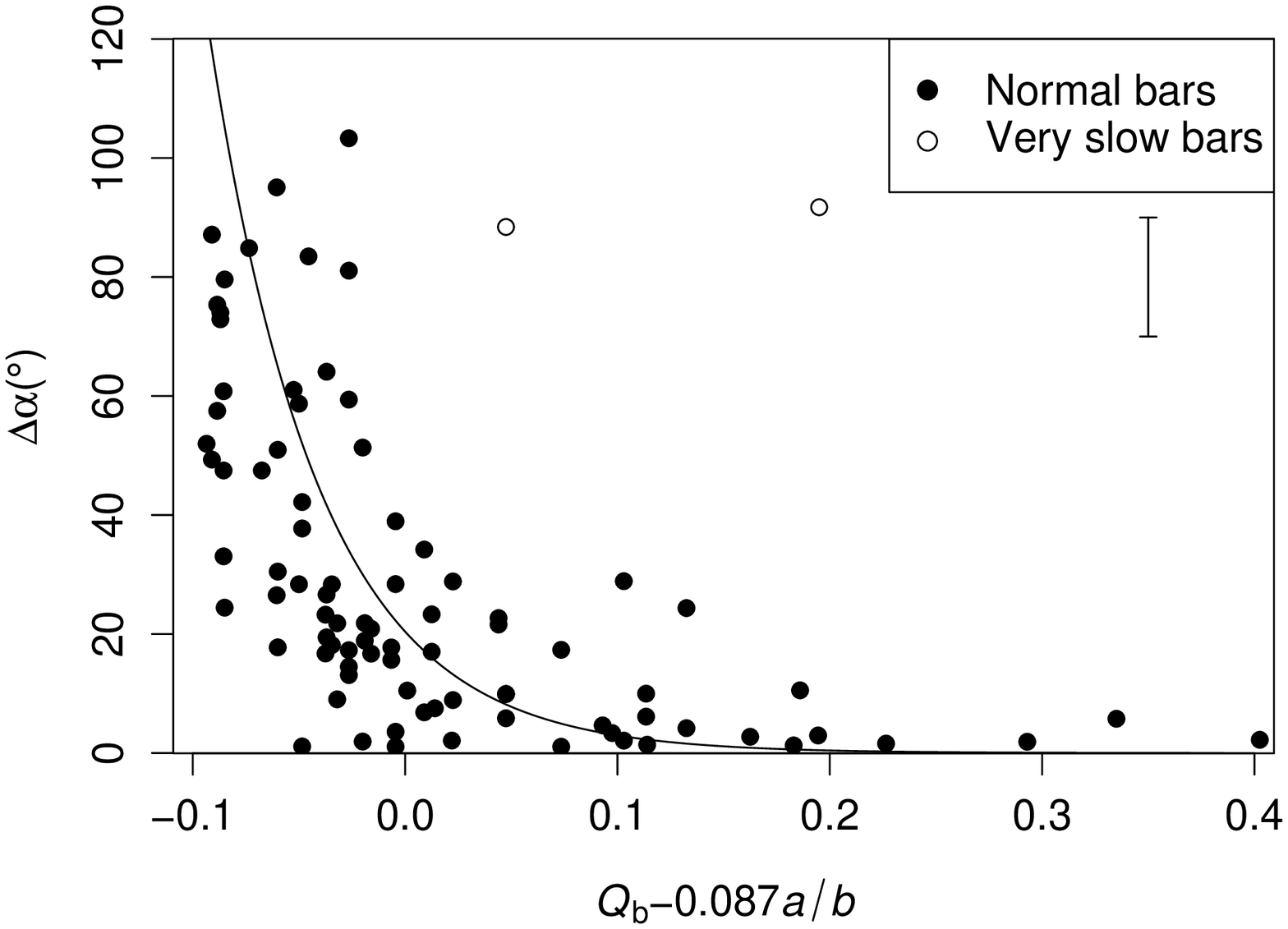}
\caption{Dust lane curvature as a function of bar strength parameter $Q_{\rm b}$ for 55 observed galaxies (left panel), and, for a set of 88 modelled galaxies, as a function of a combination $Q_{\rm b}$ and the bar ellipticity parameter $a/b$ which minimizes the spread. The curve indicates the best fit to the `normal' bars. Reproduced with permission from Comer\'on et al. (2009).}
\label{dustlanesfig}       
\end{figure}

The predicted correlation can indeed be recognised in their results, reproduced here in Fig.~\ref{dustlanesfig} (left panel). But the line which would indicate a linear relationship between stronger bars and less curved dust lanes is, in fact, only the upper envelope to the distribution of points. This is not scatter, as indicated by the typical uncertainties of each data point. The strength of the bar does not prescribe the degree of curvature that the dust lanes can have, but instead only provides an upper limit. Hence, strong bars can only have straight dust lanes, whereas weak bars allow their dust lanes to be either curved or straight. 

To investigate what factor other than the bar strength might cause the degree of curvature of the dust lanes, Comer\'on et al. (2009) analysed a set of 238 simulated galaxies, in 88 of which the dust lane curvature could be measured. The results of this, as shown in the right panel of Fig.~\ref{dustlanesfig}, show that the scatter can be greatly reduced when the bar strength is described as a linear combination of $Q_{\rm b}$ and $a/b$, the quotient of the major and minor axes of the bar. These are both bar parameters, and since no other parameters (describing, e.g., the bulge) could be identified that reduce the scatter in the original diagram, this is proof that indeed the dust lane curvature is predominantly determined by the parameters of the bar. Just how the linear combination of $Q_{\rm b}$ and $a/b$, as plotted in Fig.~\ref{dustlanesfig}, must be interpreted is an interesting question that remains to be explored further.

\subsection{Bars and spirals}

Bars also affect the disk regions outside them. This is perhaps best seen in the way they drive spiral density waves. That they do is expected from theory and numerical modelling, but also in this case it has been hard to confirm observationally. Previously, various observational links between bars and spirals had been reported (e.g.,  Kormendy \& Norman 1979; Elmegreen \& Elmegreen 1982), but more recent works have reported either good (Block et al. 2004), some (Buta et al. 2005, 2009), or no correlation (Seigar et al. 2003; Buta et al. 2009; Durbala et al. 2009) between bar and spiral strengths. The reasons for the discrepant results include sample size, methodology, and data quality, as summarised by Salo et al. (2010). The latter authors re-investigated this problem using the same data which had led Buta et al. (2005, 2009) to inconclusive results, but using a novel analysis approach.

\begin{figure}
\includegraphics[width=\textwidth]{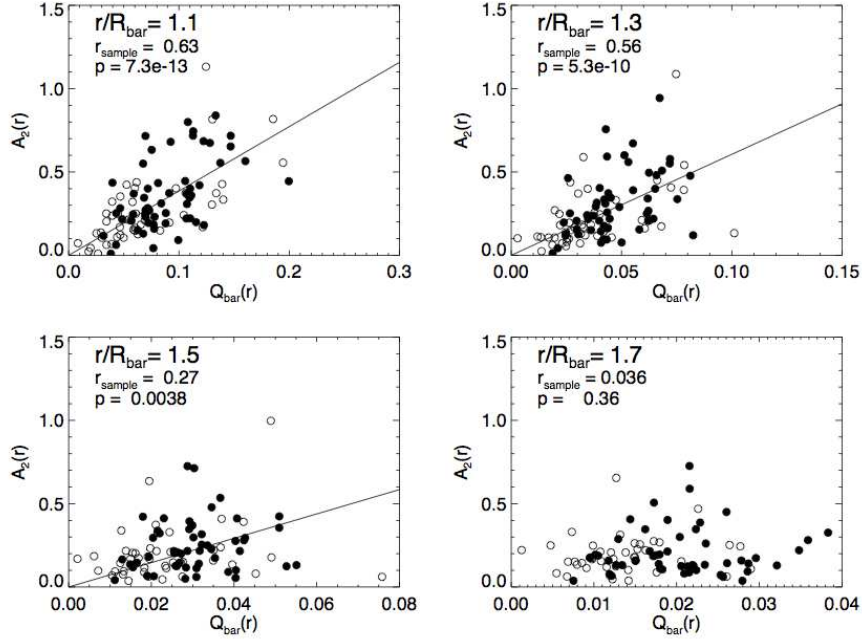}
\caption{Relation between the local bar forcing ($Q_{\rm bar}(r)$) and the local spiral amplitude ($A_2(r)$) for 103 barred galaxies, and at four distances $r/R_{\rm bar}$, normalised to the bar length. The values $r_{\rm sample}$ and $p$ indicate the correlation coefficient and significance; the best fit is indicated by a line in the three cases of highest significance. Open and filled symbols indicate short and long bars and yield the same distribution. Reproduced with permission from Salo et al. (2010).}
\label{Salofig}       
\end{figure}

Salo et al. (2010), rather than compare maxima of bar strength and spiral density amplitude, compare the {\it local} bar forcing and spiral amplitude as a function of distance. As reproduced in Fig.~\ref{Salofig}, this yields a correlation between bar forcing and spiral amplitude, strongest near the ends of the bar, but statistically significant up to one and a half times the bar radius. This confirms that the stellar spirals represent a continuation of the bar mode, or are driven by the bar through some mechanism. Outside the region of influence of the bar, the spirals may be independent modes, or transient. As the correlation between bar and spiral is similar for early- and late-type spirals, and for small and large bars, Salo et al. (2010) conclude that the forcing of the spiral by the bar is a general occurrence.

\subsection{Bars and secular evolution}

As a final illustration of the possible use of bars as tracers of secular evolution in galaxies, we cite the recent work by Buta et al. (2010a). These authors use images from the Near-Infrared S0 Survey (NIRS0S; Laurikainen et al. 2005; Buta et al. 2006; Laurikainen et al. 2006) with similar images of spirals from the literature to study the bar strength {\it distribution} in lenticular as compared to spiral galaxies. This is a pertinent question because lenticular galaxies remain somewhat of an enigma. They have been positioned between ellipticals and spirals in galaxy classification schemes since the earliest work by Hubble (1926), and one of the main questions is whether the S0 galaxies are more closely related to elliptical or spiral galaxies.

In modern Lambda Cold Dark Matter ($\Lambda$CDM) cosmology, ellipticals and the bulges of spirals are formed early in the evolution of the Universe by mergers, and their properties were thus established already early on. In this framework, S0s may either be formed as ellipticals, or alternatively they may be transformed spirals, formed as the disks have lost their gas by some stripping mechanism. In a recent paper also based on NIRS0S images, Laurikainen et al. (2010) confirm from photometric scaling relations that the formative processes of bulges and disks in S0s are coupled, and that the bulges of S0s are similar to those of spirals with bright bulges. They conclude that spiral galaxies with bulges brighter than $M_K{\rm (bulge)}<-20$\,mag can evolve directly into S0s, due to stripping of gas followed by truncated star formation. This is {\it prima facie} evidence for secular evolution.

\begin{figure}
\sidecaption
\includegraphics[scale=.3]{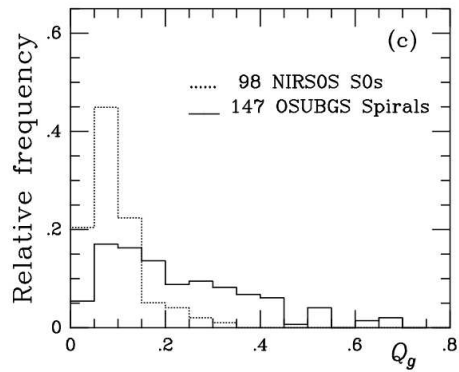}
\caption{Histograms of the bar strength distribution in S0 and spiral galaxies. Reproduced with permission from Buta et al. (2010a).}
\label{Butabarfig}       
\end{figure}

The bar strength distribution of Buta et al. (2010a) shows (Fig.~\ref{Butabarfig}) that S0 galaxies on average have weaker bars than spiral galaxies in general, and even than early-type spirals, of type S0/a and Sa. Several studies, mostly based on contrast, have shown that bars in early-type galaxies are stronger and longer than those in later-type galaxies (e.g., Elmegreen \& Elmegreen 1985), other studies report weaker bars in lenticulars as compared to spirals (e.g., Aguerri et al. 2009, who used a strength measure based on bar ellipticity). When considering the gravitational forcing of the bar, as measured, e.g., through the $Q_{\rm b}$ parameter, the non-axisymmetric bar forcing is to a large extent cancelled in early-type galaxies by the axisymmetric contribution to the potential of their more massive bulges (e.g., Laurikainen et al. 2004b), so that in early-type galaxies bars are longer and have larger $A2$-amplitudes, but have lower $Q_{\rm g}$ and ellipticity than spiral galaxies (Laurikainen et al. 2007). 

The differences found by Buta et al. (2010a) are significant, and only partly due to the dilution of the bar torques by the large bulges of the S0s, or the thicker disks in S0s. They tell us that if indeed S0s have evolved from spirals, the bar evolution must have continued after the gas depletion, which might also be suggested by the lower frequency of bars (Laurikainen et al. 2009) and the higher frequency of lenses (Kormendy 1979; Laurikainen et al. 2007) in S0s as compared to spirals. Buta et al. (2010a) speculate that the bars in early-type galaxies can be slightly skewed, so that a potential-density phase shift (Zhang \& Buta 2007) can evolve the stellar distribution, leading to continued bulge building and bar weakening.

\section{Rings}
\label{rings}

Rings in galaxies are common, and are thought to trace resonances in the disk. This means that they can be used as easily observable tracers of the underlying dynamical structure of the galaxy. Of the three main categories of `resonance rings', outer, inner, and nuclear rings, the former two types can be recognised on standard optical imaging of galaxies, and have thus been classified as part of the main galaxy catalogues. In particular, the RC3 catalogue (de Vaucouleurs et al. 1991) assigns the categories `R' for outer ring, and `r' for inner ring (pseudorings were also classified but are outside the scope of this review). The classification scheme has been perfected in the De Vaucouleurs Atlas of Galaxies (Buta et al. 2007), and most recently by Buta et al. (2010b). In the latter paper, we used {\it Spitzer} mid-IR images from the S$^4$G survey to classify 207 galaxies, and introduced also the third class of rings, nuclear rings, into the classification (`nr'). 

In contrast to inner and outer rings, it is much harder to recognize nuclear rings in the images typically used for galaxy classification. As a result, there has not been a complete inventory of nuclear rings. This we have remedied with AINUR (Comer\'on et al. 2010), where 113 {\it bona fide} nuclear rings in 107 galaxies were catalogued and studied. The main aims of AINUR were, first, to make an inventory as complete as possible of nuclear rings in the local Universe, and second, to study in a statistical sense their properties in relation to the properties of their host galaxies. The importance of nuclear rings lies primarily in the fact that they are assumed to be tracers of recent gas inflow to the circumnuclear region (Combes \& Gerin 1985; Shlosman et al. 1990; Athanassoula 1994; Knapen et al. 1995; Heller \& Shlosman 1996; Combes 2001; Regan \& Teuben 2003). They form significant quantities of stars (Kennicutt et al. 2005) and may thus help the secular building of a bulge, and are close to the region where non-stellar activity occurs in many galaxies (Knapen 2005). 

\begin{figure}
\sidecaption
\includegraphics[width=0.6\textwidth]{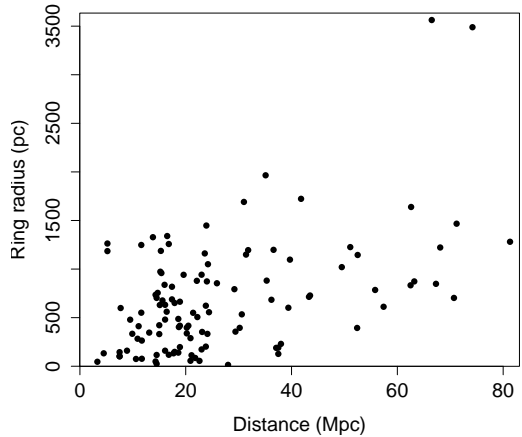}
\caption{Distribution of nuclear ring radius as a function of distance to the host galaxy for the 113 nuclear rings in AINUR. Reproduced with permission from Comer\'on et al. (2010).}
\label{AINUR_f1}       
\end{figure}

As seen in Fig.~\ref{AINUR_f1}, the radii of nuclear rings vary from a few tens of pc (the limit allowed by space-based imaging) to over 3\,kpc. We can reliably recognise nuclear rings in galaxies up to a distance of some 80\,Mpc, although only the larger rings are seen in galaxies over some 20\,Mpc away. In AINUR, we define a nuclear ring primarily as a star-forming ring-like feature in the proximity of the nucleus, employing a number of additional criteria relating to the ring width and radius. It is usually easy to decide whether rings are nuclear or inner, because inner rings occur near the end of a bar (Schwarz 1984), whereas nuclear rings occur well within the bar. In the absence of a bar, and if only one ring is present, distinguishing between nuclear and inner rings may be impossible. We do not include pseudo-rings (intermediate between nuclear spiral and nuclear ring) but the dividing line is somewhat fuzzy.

AINUR catalogues 113 nuclear rings in a total of 107 galaxies, 18 of which are unbarred and 78 of which are barred disk galaxies. Most of these rings had been reported before and were confirmed from archival {\it Hubble Space Telescope} ({\it HST}) or other imaging (a significant number of other features called nuclear rings in the literature could not be confirmed and were not included; this category includes inner rings, pseudo-rings, or galaxies for which no high-quality imaging is available). A total of 17 previously unreported nuclear rings were discovered. The AINUR catalogue can be considered a complete list of all {\it bona fide} nuclear rings known at present, but as more high-quality (mainly {\it HST}) imaging comes available, it is likely that a few more nuclear rings are discovered. 

On the basis of a complete sample of galaxies within AINUR, the fraction of disk galaxies with morphological types in the range $-3<T<7$ that host a star-forming nuclear ring is confirmed to be 20\%$\pm$2\% (see also Knapen 2005). This is a high fraction, and considering that the star formation activity is mostly seen in emission from massive stars (UV, H$\alpha$) and thus quite possibly short-lived (Allard et al. 2006; Sarzi et al. 2007), one can conclude that nuclear rings are very frequent indeed.

\begin{figure}
\includegraphics[width=0.5\textwidth]{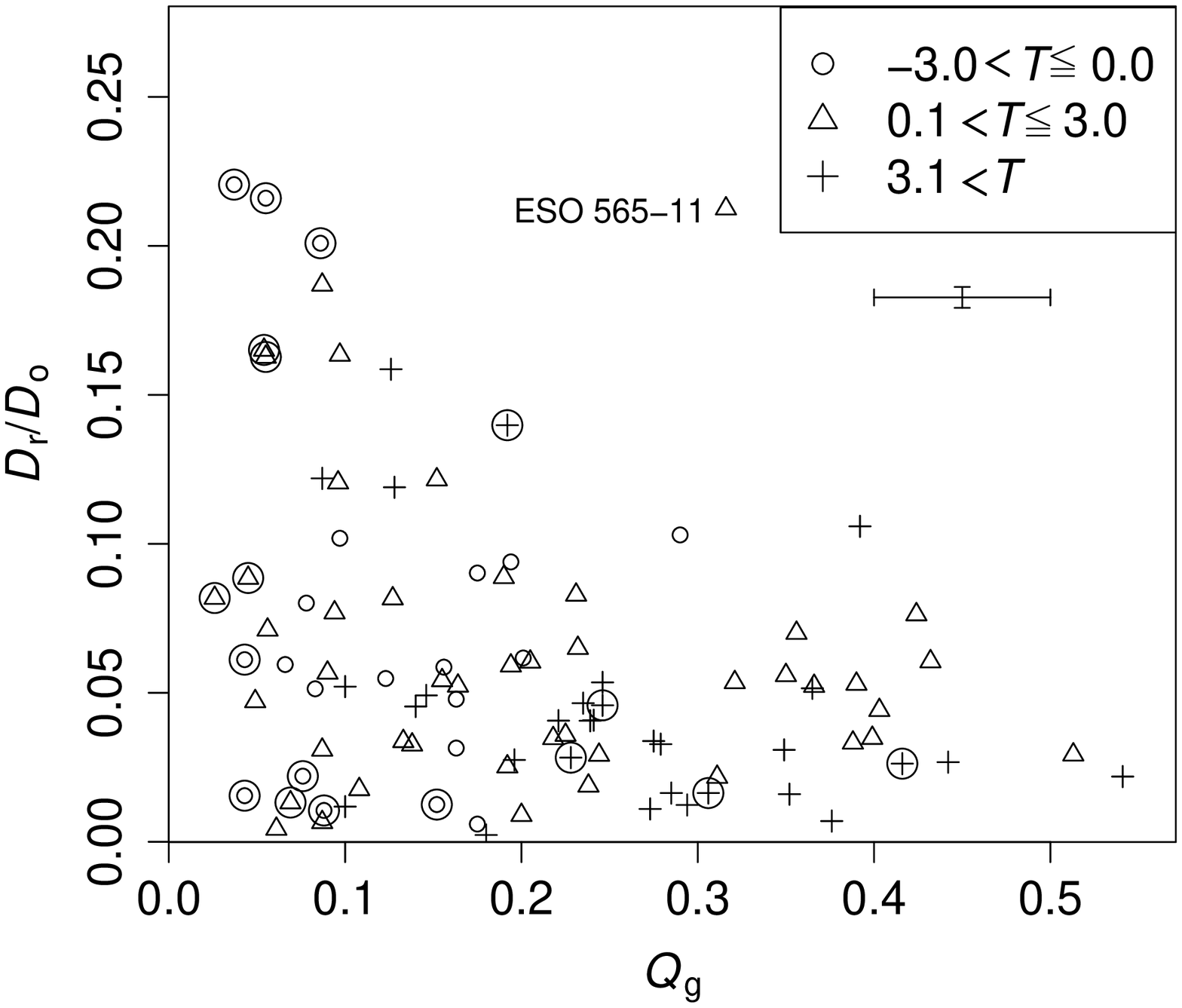}
\includegraphics[width=0.5\textwidth]{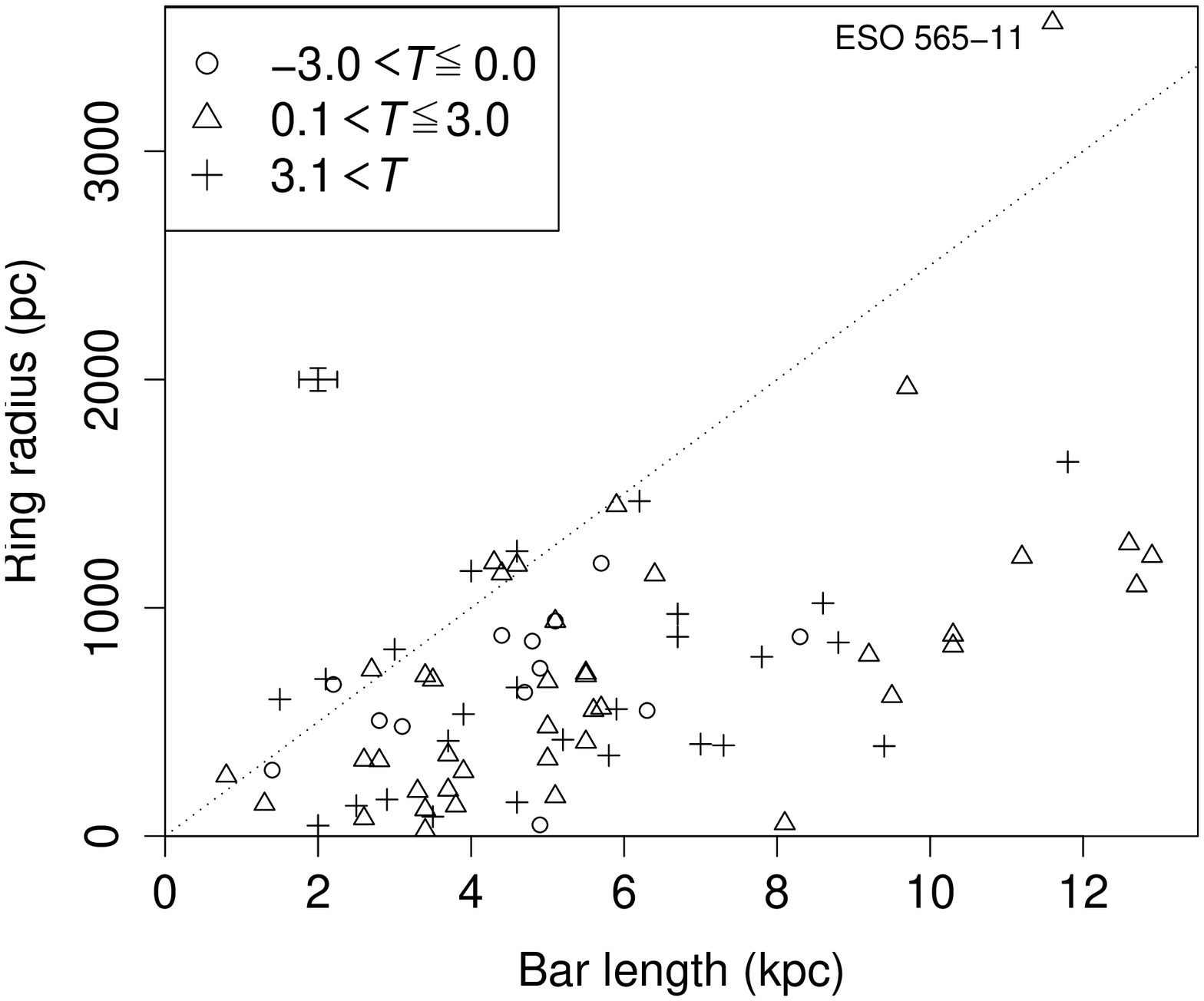}
\caption{Left panel: relative nuclear ring size, normalised by the host galaxy disk size, as a function of the bar strength parameter $Q_{\rm g}$, and separated by morphological type. Encircled symbols denote unbarred galaxies. Right panel: Absolute nuclear ring size versus the bar length. Dashed line indicates bar lengths four times the ring radii. Reproduced with permission from Comer\'on et al. (2010).}
\label{AINUR_f56}       
\end{figure}

Using basic nuclear ring parameters such as size determined from the AINUR imaging (mostly {\it HST}), and host galaxy and bar parameters (e.g., bar length, bar ellipticity, $Q_{\rm g}$) determined from near-IR images from the 2 Micron All-Sky Survey (2MASS), Comer\'on et al. (2010) explored relations between the nuclear rings and their host galaxies and bars (where present).  

Fig.~\ref{AINUR_f56} shows two examples. In the left panel, it is seen that for small values of $Q_{\rm g}$ (the non-axisymmetric torque of the galaxy, which in most cases is dominated by the bar---small $Q_{\rm g}$ values thus generally denote weak bars, and high values strong bars) a wide range of relative nuclear ring sizes is allowed, where galaxies with high $Q_{\rm g}$ can only have small rings. This had been found from a much smaller sample of nuclear rings by Knapen (2005), and seen in simulations by Salo et al. (1999). Large nuclear rings can only occur in weak bars. Analogous to what we found for the bar dust lanes (Sect.~2.2), the bar strength does not prescribe the nuclear ring size, but does set an upper limit. What determines the exact size of a nuclear ring within the range allowed by its bar is not clear, but must be related to the shape of the gravitational potential which conditions  the location of the Inner Lindblad Resonances (ILRs).

The right panel of Fig.~\ref{AINUR_f56} shows how the length of the bar limits the radius of the nuclear ring (to around $r_{\rm ring}=r_{\rm bar}/4$), while rings smaller than the upper limit are apparently allowed. Both these relations confirm that the size of the nuclear ring is limited by basic parameters of the bar, and thus that the rings must be closely related dynamically to the bar.

\begin{figure}
\includegraphics[width=\textwidth]{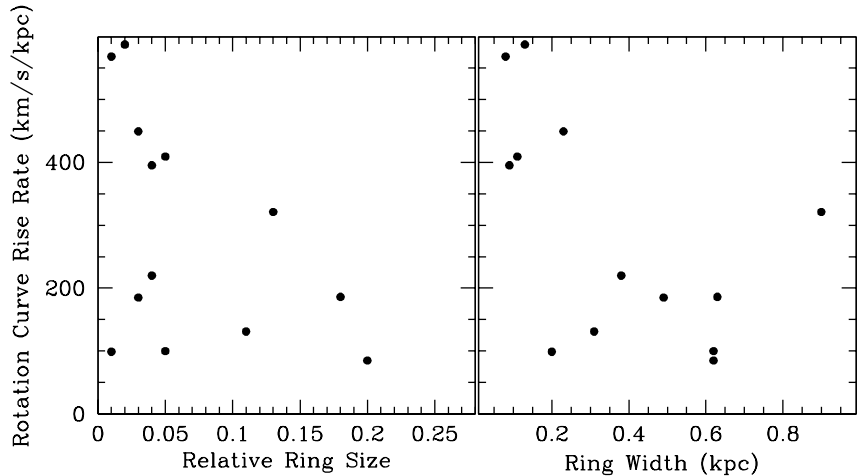}
\caption{Rise rate of the inner part of the rotation curve as a function of the nuclear ring size relative to the disk size (left panel), and the nuclear ring width (right panel). The thirteen different galaxies in this sample are indicated with different symbols. Data from Mazzuca et al. (2010).}
\label{Mazz_1}       
\end{figure}

Further evidence for the intricate links between host galaxy, bar, and nuclear ring is provided by Mazzuca et al. (2010) from a study of the inner rotation curves of 13 nuclear ring host galaxies. They obtained H$\alpha$ velocity fields using the DensePak integral field unit on the 3.5\,m Wisconsin, Indiana, Yale \& NOAO (WIYN) telescope and the TAURUS Fabry-Perot instrument on the 4.2\,m William Herschel Telescope. From these, they derived rotation curves after assuring that the non-circular motions were small enough to allow such a derivation. As the rotation curve is, theoretically, intricately linked to the gravitational potential and to the location of the ILRs which give rise to the nuclear rings (Knapen et al. 1995), Mazzuca et al. (2010) then parametrized the inner part of the rotation curve and compared the results with the location of the nuclear ring, as well as the properties of the bar.

The main rotation curve parameter considered by Mazzuca et al. (2010) is its rise rate, defined as the ratio between the velocity difference between that at the turnover point (where the rotation curve stops rising and flattens out) and that at the origin, and the difference in radius between those two points. As can be seen in Fig.~\ref{Mazz_1}, plotting the rise rate of the rotation curve as a function of the relative nuclear ring size or the nuclear ring width yields diagrams reminiscent of the ones we have presented earlier in this paper: they outline an upper limit, indicating that ring size and width limit the rotation curve rise rate or vice versa. Large nuclear rings, and wide nuclear rings, can only occur when the rotation curve rises slowly. In the case of a rapidly rising rotation curve, the nuclear ring can only be small, and narrow.

Mazzuca et al. (2010) thus present a neat observational confirmation that the rotation curve and the metric parameters of the nuclear ring are intricately linked. The physical background of this link must include the location of the resonances: linear theory\footnote{In the linear approximation the gravitational potential of the bar is considered to be axisymmetric. While this may be reasonable for weak bars, it is not correct in the case of a strong bar (Sellwood \& Wilkinson 1993; Shlosman 2001). Nevertheless, as an intuitive tool it can provide a useful illustration.} predicts that ILRs, and nuclear rings, occur where the rate of change of the circular velocity is highest---this is where the rotation curve turns over. The rise rate of the rotation curve is thus a measure of where the ILRs are located: as rotation curves tend to flatten out at circular velocities of between 100 and 200\,km\,s$^{-1}$, a high rise rate means that the ILRs are close to the nucleus, and thus that the nuclear ring must be small. It also implies that the ILRs are close together radially, which leads to narrower nuclear rings. All this is exactly as seen observationally. A slowly rising rotation curve implies that the ILRs can be located further out from the nucleus, but also that the distance between the inner and outer ILRs (which in linear theory limit the radial range where the nuclear ring can occur) can increase. Depending on the precise shape of the gravitational potential in the inner region, the inner ILR can still be relatively close to the nucleus, which might explain that in galaxies with slowly rising rotation curves large and wide nuclear rings can occur, but small and narrow ones are not excluded.

\begin{figure}
\sidecaption
\includegraphics[width=0.6\textwidth]{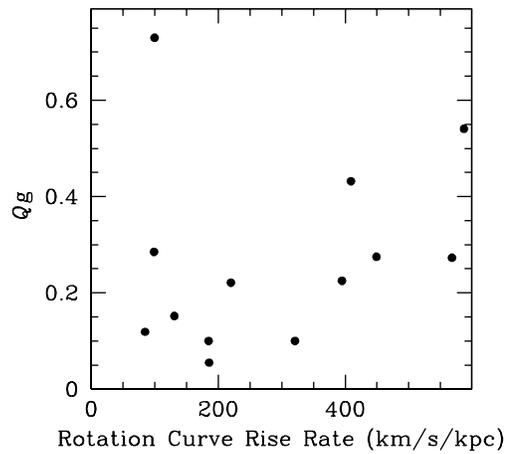}
\caption{Non-axisymmetric torque parameter $Q_{\rm g}$, which can be interpreted as an indicator of the strength of the bar, plotted against the rotation curve rise rate. Data from Mazzuca et al. (2010).}
\label{Mazz_2}       
\end{figure}

Fig.~\ref{Mazz_2} shows that the rotation curve rise rate is also located to the bar strength, measured here through the non-axisymmetric torque parameter $Q_{\rm g}$. The figure indicates a general trend of more steeply rising rotation curves as $Q_{\rm g}$ increases (the one notable exception is NGC~1530 which is a galaxy with a bar that is exceptional in many more ways). It is not clear whether the two parameters plotted here are directly or indirectly related---we have already seen that stronger bars tend to host smaller rings. But as a general conclusion, it is beyond doubt now that the underlying dynamics of a host galaxy influences its appearance and kinematics, as well as detailed aspects of some of the structural components such as bars and rings.

\section{Future work: the S$^4$G survey}
\label{future}

The future of the field of secular evolution in galaxies is in no small measure infrared. The {\it Herschel} satellite has just started producing spectacular data in nearby galaxies, the {\it James Webb Space Telescope} ({\it JWST)} is being prepared for launch, and the {\it Spitzer Space Telescope} ({\it SST}; Werner et al. 2004) is still producing wonderful data, even now its coolant has been depleted and the telescope is `warm'. 

The {\it Spitzer} Survey of Stellar Structure in Galaxies (S$^4$G; Sheth et al. 2010) is designed to be the definitive survey of the distribution of stellar structure in the nearby Universe. Over the two years of the warm mission of the {\it SST}, the S$^4$G will observe the stellar mass distribution in a volume-, magnitude- and size-limited ($d<40$\,Mpc, $m_B < 15.5$, $D_{25} > 1$\,arcmin) sample of 2,331 galaxies using the Infrared Array Camera (IRAC; Fazio et al. 2004) at 3.6 and 4.5\,micron. This survey will provide an unprecedented set of imaging data which will be used to study the stellar structure in galaxies. Among many other topics, the S$^4$G data will allow the study of how outer disks and halos are formed, how the formation and evolution of galactic structures are affected by galaxy interactions, or which structural parameters govern secular galaxy evolution. The large sample, ranging from dwarfs to spirals to ellipticals will allow for such structural studies both as a function of stellar mass and as a function of environment, vital to test cosmological simulations predicting the mass properties of present day galaxies.

The S$^4$G survey was awarded 637.2 hours of time on the Spitzer telescope, and the observations are in progress. It is expected that the survey will be completed by mid-2011. The data are being analysed using dedicated pipeline software mostly developed by the S$^4$G consortium, which will perform a basic reduction, sky subtraction and mosaicing of the images, will produce masks of foreground stars, will measure basic parameters of all galaxies, and will deliver multi-component decompositions (see Sheth et al. 2010 for a detailed description). 

\begin{figure}
\includegraphics[width=\textwidth]{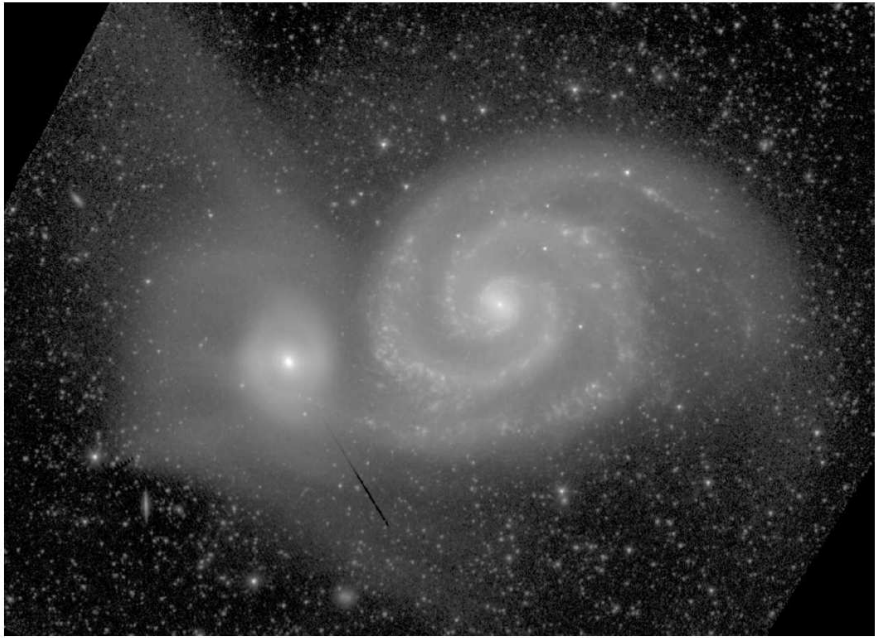}
\caption{{\it Spitzer} 3.6\,$\mu$m S$^4$G image of the galaxy M51 (NGC~5194, right) and its companion NGC~5195. From Buta et al. (2010b).}
\label{s4gfig}       
\end{figure}

As an example of the data quality, including depth of imaging and field of view related to the size of the galaxy,  that will be delivered by the S$^4$G for {\it all} survey galaxies,  we present in Fig.~\ref{s4gfig} a 3.6\,$\mu$m image of the well-known galaxy M51. The image is reproduced from a paper by Buta et al. (2010b), which discusses the morphology of the first 207 S$^4$G galaxies. Not only does the image show the intricate outer tidal structure caused by the interaction of M51 and its companion NGC~5194 (see Salo \& Laurikainen 2000), it also illustrates how a high-quality dust-free vision of a galaxy can lead to a better insight into its nature.  On the basis of this new image, Buta et al. (2010b) were able to revise the morphological classification of M51 from SA(s)bc pec (RC3) to S\underline{A}B(rs,nr)bc, but much more spectacular is the revision of the type of the companion of M51, NGC~5195, which was classified as an I0 pec galaxy in de Vaucouleurs et al. (1991), but which Buta et al. (2010b) could classify as SAB(r)0/a pec. The bar and ring giving rise to this new classification can be easily seen in Fig.~\ref{s4gfig}.

The combination of deep, wide-field, mid-infrared imaging and a large sample offered by the S$^4$G will allow significant progress on most of the issues discussed earlier in this paper. For instance, the survey will allow a comprehensive and definitive study of bar and spiral arm properties across a wide range of host galaxy parameters. The new images will allow a more complete study of the host galaxies of not only nuclear, but also inner and outer rings, including the interesting cases of unbarred ring hosts. The combination of the {\it Spitzer} images with ancillary data, for instance {\it GALEX} UV imaging, H$\alpha$ imaging, atomic and molecular gas observations, and kinematic observations offers almost limitless opportunities for scientific progress. The S$^4$G should thus lead to a much more complete understanding of galaxy evolution, and of the precise role of secular evolution. 

\section{Conclusions}
\label{conclusions}

This review has highlighted some recent work which further illuminates the tight physical links between bars, nuclear rings, and their host galaxies. These relationships are shedding light on the underlying dynamical structure, and on the overall process of secular evolution of galaxies. The dynamics of bars is now rather well understood. They indeed stimulate inflow of gaseous material from the disk to the central region of a galaxy, and thus inflowing gas can help build the bulge and drive evolution, even across morphological types. Nuclear rings are common, and transform significant amounts of gas into stars, again increasing the bulge mass. Nuclear rings occur mostly in barred galaxies, but the fraction occurring in unbarred galaxies is similar to the overall fraction of unbarred disk galaxies, so a causal link is not proven. Ovals, interaction event, or possibly even strong spiral arms can cause enough non-axisymmetry in the gravitational potential to stimulate the formation of a nuclear ring, even in the absence of a bar. This shows how ubiquitously secular evolution in galaxies can occur.

The future for this area of research is bright, because significant amounts of new data are being collected which will allow us to increase sample sizes and progress to study secular evolution on a more fundamental basis. We explicitly mentioned the S$^4$G survey, which will deliver deep, mid-IR imaging covering the complete disks of more than 2300 galaxies in the local Universe. 

{\it Acknowledgements} I thank my co-workers Ron Buta, S\'ebastien Comer\'on, Eija Laurikainen, and Heikki Salo for comments on an earlier draft of this paper. I thank them and my other collaborators, including those on the S$^4$G team, for stimulating discussions and scientific progress. 

%
%
%

\end{document}